\documentclass[prl,twocolumn,showpacs,superscriptaddress,notitlepage]{revtex4-2}
\usepackage{makeidx}
\usepackage[utf8]{inputenc}
\usepackage{graphicx}
\usepackage{amsmath}
\usepackage{mathrsfs}
\usepackage{graphicx}
\usepackage{braket}
\usepackage[subrefformat=parens,labelformat=parens,caption=false]{subfig}
\usepackage{amsfonts}
\usepackage{dcolumn}
\usepackage{bm}
\usepackage{bbm}
\usepackage{color}
\usepackage[dvipsnames]{xcolor}
\usepackage{esint}
\usepackage{lineno}
\usepackage{verbatim}
\usepackage{cancel}
\usepackage[breaklinks,
            colorlinks,
            urlcolor=blue,
            linkcolor=blue,
            anchorcolor=blue,
            citecolor=blue]{hyperref}        
\usepackage{soul}
\usepackage{amssymb}

\newcommand{\change}[1]{{\color{black} #1}}

\definecolor{THc}{rgb}{0.2,0.9,0.5}

\begin{document}
\title{\change{Ground-state selection via many-body superradiant decay }}
\begin{abstract}
\change{For a single particle, relaxation into different ground states is governed by fixed branching ratios determined by the transition matrix element and the environment. Here, we show that in many-body open quantum systems the occupation probability of one ground state can be boosted well beyond what is dictated by single-particle branching ratios.} Despite the competition, interactions suppress all but the dominant decay transition, leading to a ‘winner takes all’ dynamic where the system primarily settles into the dominant ground state. We prove that, in the presence of permutation symmetry, this problem is exactly solvable for any number of competing channels. Additionally, we develop an approximate model for the dynamics by mapping the evolution onto a fluid continuity equation, and analytically demonstrate that the dominant transition ratio converges to unity as a power law with increasing system size, for any branching ratios. This near-deterministic preparation of the dominant ground state has broad applicability.  As an example, we discuss a protocol for molecular photoassociation where collective dynamics effectively acts as a catalyst, amplifying the yield in a specific final state. Our results open new avenues for many-body strategies in the preparation and control of quantum systems.

\end{abstract}
\author{Wai-Keong Mok}
\email{darielmok@caltech.edu}
\affiliation{Institute for Quantum Information and Matter, California Institute of Technology, Pasadena, California 91125, USA} 
\author{Stuart J. Masson}
\affiliation{Department of Physics, Columbia University, New York, New York 10027, USA}
\author{Dan M. Stamper-Kurn}
\affiliation{Department of Physics, University of California, Berkeley, California 94720}
\affiliation{Challenge Institute for Quantum Computation, University of California, Berkeley, California 94720}
\affiliation{Materials Sciences Division, Lawrence Berkeley National Laboratory, Berkeley, California 94720}
\author{Tanya Zelevinsky}
\affiliation{Department of Physics, Columbia University, New York, New York 10027, USA}
\author{Ana Asenjo-Garcia}
\email{ana.asenjo@columbia.edu}
\affiliation{Department of Physics, Columbia University, New York, New York 10027, USA}
\maketitle

\begin{figure*}
\centering
\subfloat{%
\includegraphics[width=1.0\linewidth]{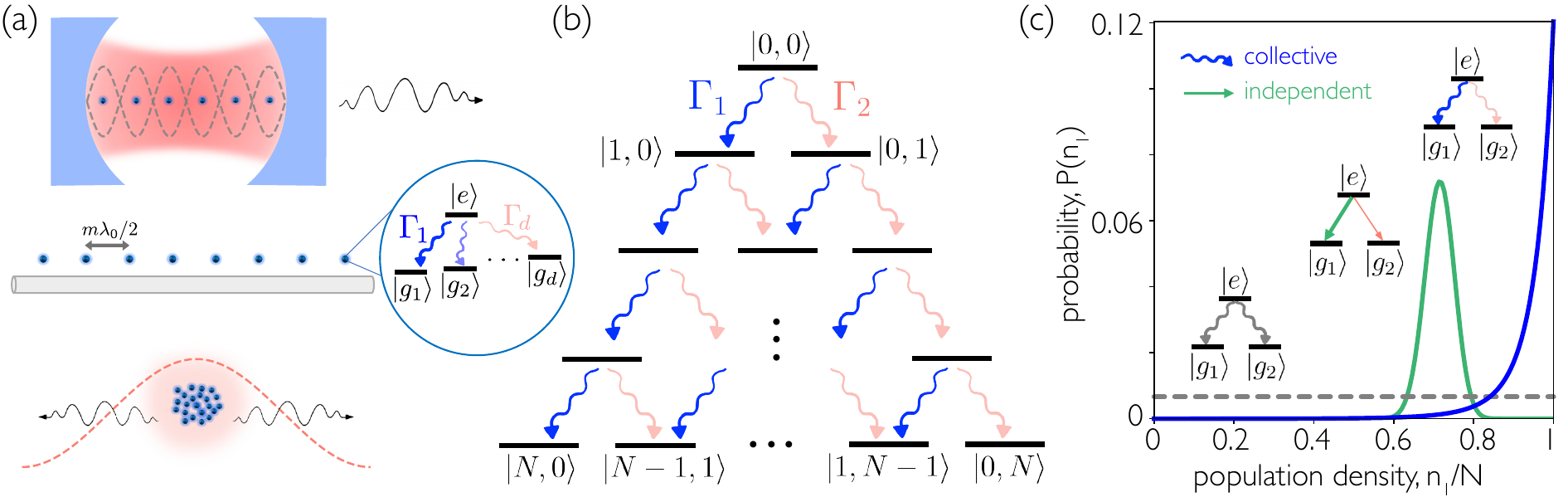}%
  \label{}%
}
\caption{\change{Ground-state selection} due to collective decay in multilevel systems. (a) $N$ multilevel emitters with $d$ ground states and a single excited state decay collectively to an environment (at rates $\Gamma_1,\ldots,\Gamma_d$), such as a ``bad'' cavity, a waveguide in the ``mirror configuration'' (where the relative distance between the emitters is a half-integer multiple of the resonance wavelength $\lambda_0$) or free space (for the latter, a dense ensemble of subwavelength volume is required to preserve permutational symmetry). (b) Dissipative dynamics can be modeled as a random walk between permutationally symmetric ground states ($d = 2$ depicted), labelled by $\ket{n_1,\ldots,n_d}$, where $n_\mu$ denotes the population of the ground state $\ket{g_\mu}$. (c) Marginal probability distribution $P(n_1)=\sum_{n_2} P(n_1, n_2)$ for the population density of the ground state $\ket{g_1}$, for $N = 150$ and $d=2$ in the steady state, after complete depletion of the fully-inverted initial state. The distributions for collective decay with $r_2 = 1$ (gray), and collective decay with $r_2 = 0.5$ (blue) are obtained via numerical simulations of Eq.~\eqref{eq:rate_eq}. The binomial distribution is plotted for independent decay with $r_2 = 0.5$ (green).}
 \label{fig:scheme}
\end{figure*}

The suppression of decay pathways into undesired quantum states is crucial for the control and manipulation of open quantum systems. Simplified theoretical models often rely on `closed' transitions where the excited state decays predominantly to a specific ground state. However, in practice, real-world emitters rarely adhere to the idealized paradigm of two-level systems. For instance, highly excited atomic states (such as Rydberg states) have many lower-energy states accessible by spontaneous emission~\cite{cong}. In photochemistry, the decay to a single ground state is often inefficient due to numerous competing pathways arising from electronic, vibrational, and rotational degrees of freedom~\cite{carr2009cold,bohn2017cold}. Solid-state emitters (such as color centers or dye molecules) also suffer from parasitic decay from phonon sidebands~\cite{atature2018material,lange2024superradiant}. Achieving closed transitions in experiments is challenging and often involves isolating two-level systems from more complex internal structures or employing repumping techniques to redirect population back into the desired states. However, these approaches are inherently limited by the natural single-particle branching ratios of the transitions. 

The natural branching ratios and dynamics of decay can, in fact, be modified by engineering the quantum system and its environment. One common approach is to tailor the dielectric environment by placing emitters within optical cavities~\cite{walther2006cavity}, waveguides~\cite{vetsch2010optical,solano2017superradiance}, or other photonic structures~\cite{goban2015superradiance,zhou2023trapped} such that the desired decay channel is Purcell-enhanced. Numerical studies with multilevel atoms~\cite{stroud_1982,sutherland2017superradiance,orioli2022emergent,masson2024dicke} and molecules~\cite{pupillo} have suggested collective emission as an alternative to circumvent limitations from single-particle branching ratios. These proposals rely on many-body, transient superradiance~\cite{dicke1954coherence,Gross1982superradiance}, a phenomenon characterized by avalanche-like behavior~\cite{clemens2004shot,masson2022universality,cardenas-lopez2023many} where decay into a given ground state enhances the probability of subsequent emission into that same state. This process effectively steers the emitters towards a specific ground state. A comprehensive analytical treatment of this physics remains lacking, which is critical for understanding and exploiting its potential.

We term the phenomenon in which correlated decay suppresses all but the most dominant emission path as \textit{ground-state selection}. As shown in Fig.~\ref{fig:scheme}(a), we consider an ensemble of emitters coupled to a reservoir. The emitters have multiple decay channels, each leading to different final states. Each decay channel can be collectively enhanced by many-body correlations that emerge dynamically, leading to competition between them, \change{effectively quenching the subdominant channels}. We find approximate steady-state solutions for the populations of the different ground states by modeling the quantum dynamics through a continuity equation for a fluid. We prove that, despite the competition, the population density of the dominant ground state exhibits a power-law convergence to unity for any branching ratio, with the power-law exponent characterized by the ratio between dominant and subdominant decay rates. This is supported by a rigorous analysis of the exact steady-state solution. \change{Our analytical treatment provides novel physical insights on the phenomenon beyond the numerical observations in Refs.~\cite{sutherland2017superradiance,orioli2022emergent,masson2024dicke}.} We apply our framework to the problem of photochemistry, where suboptimal branching ratios limit the effectiveness of molecule creation, direct laser cooling~\cite{TarbuttFitchAAMOP21_LaserCooledMolecles}, and optical imaging. Specifically, in molecular photoassociation of strontium dimers, we demonstrate that \change{ground-state selection} greatly enhances sample purity.

We consider an (undriven) ensemble of $N$ identical emitters. \change{Each emitter has} a level structure consisting of a single excited state $\ket{e}_{i}$ and $d$ ground states labelled $\ket{g_{\mu}}_i$, with $\mu \in \{1,\ldots,d\}$ and $i \in \{ 1,\ldots,N\}$. \change{The level structure describes natural emitters including atoms, trapped ions, and color centers.} The emitters are symmetrically coupled to a Markovian environment, with separate couplings for each transition $\ket{e}\to\ket{g_\mu}$. Physically, this can be realized by coupling the emitters to near-resonant cavity modes (in the ``bad cavity'' or ``weak coupling'' limit, and placing the emitters in locations where they couple with the same strength to the cavity mode), to a single-mode waveguide in the ``mirror configuration'' (where the emitters are separated by a half-integer multiple of the wavelength~\cite{chang_2012,polzik}),  or via free-space interactions in a dense ensemble of subwavelength volume, as depicted in Fig.~\ref{fig:scheme}(a). We assume that photons from different transitions can be resolved either by polarization or frequency (for the waveguide configuration, we assume that the transition frequencies are approximately equal to fulfil the mirror condition). The system dynamics is modelled by the master equation 
\begin{equation}
    \dot{\rho} = -i\left[\sum_{\mu=1}^{d} \chi_\mu \hat{A}_\mu^{\dag} \hat{A}_\mu,\rho \right] + \sum_{\mu=1}^{d} \Gamma_\mu \mathcal{D}[\hat{A}_\mu]\rho,
\label{eq:ME}
\end{equation}
where $\chi_\mu$ are the coherent interaction rates, $\hat{A}_\mu = \sum_{i=1}^{N} (\ket{g_\mu}\bra{e})_i$ is the collective lowering operator on the transition $\ket{e}\rightarrow\ket{g_\mu}$, and $\mathcal{D}[\hat{A}]\rho \equiv \hat{A} \rho \hat{A}^\dag - \{\hat{A}^\dag \hat{A},\rho\}/2$. Collective dissipation occurs with rates $\Gamma_{\mu}$ (identical to the single-emitter ones), and we assume the ordering $\Gamma_1 \geq \Gamma_2 \geq \ldots \geq \Gamma_d$, such that $\Gamma_1$ is the dominant decay rate. For convenience, we denote the ratio between dominant and subdominant decay rates as $r_\mu \equiv \Gamma_\mu/\Gamma_1$. \change{Our model also encompasses a more general scenario where some of the transitions are indistinguishable, such that $d$ is the number of distinguishable channels.}

\begin{figure*}
\centering
\subfloat{%
\includegraphics[width=\linewidth]{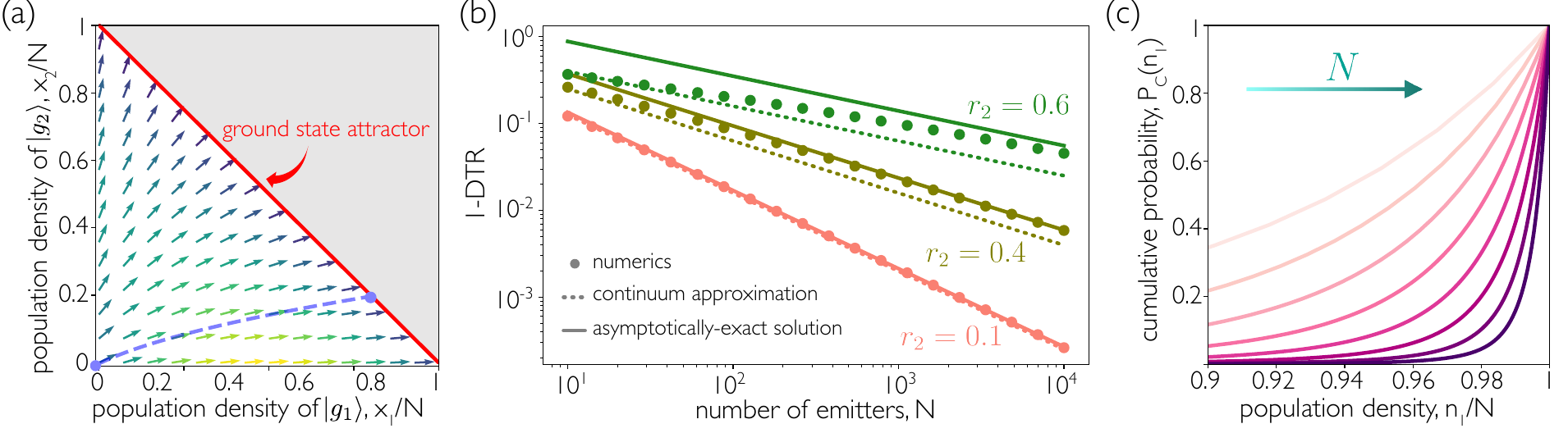}%
  \label{}%
}
\caption{(a) Velocity vector field of the continuum model for $r_2 = 0.5$ and $N=10$. The dashed line represents the flow trajectory in the single-particle approximation starting from the fully excited state $(x_1,x_2)=(0,0)$, where $x_{1(2)}$ denotes the population of $\ket{g_{1(2)}}$. Darker color indicates lower velocity. The red solid line shows the neutrally stable ground state attractor. The unphysical region (where the ground state population is larger than $N$) is shown in gray.  (b) Population density outside the dominant ground state [$1-\text{Dominant transition ratio (DTR)}$] against number of emitters $N$, for various decay ratios $r_2=\Gamma_2/\Gamma_1$. The points are obtained from numerically solving the rate equation~\eqref{eq:rate_eq}, while the dotted lines denote the approximate formula~\eqref{eq:DTR_approx}. Solid lines denote the exact asymptotic solution~\eqref{eq:DTR_conservation}. (c) Cumulative distribution (i.e., total probability of the dominant ground state population being at most $n_1$) $P_c(n_1) = \sum_{n_1^\prime=0}^{n_1} P(n_1^\prime)$. Data obtained from numerical simulation of the rate equation~\eqref{eq:rate_eq}, with $r_2 = 0.4$. Darker colors indicate larger number of emitters, with $61 \leq N \leq 10^4$. All plots are made for $d=2$ ground states.}
 \label{fig:power_law}
\end{figure*}

Dynamical evolution can be understood as a random walk in the subspace of permutationally symmetric states, as shown in Fig.~\ref{fig:scheme}(b). Since Eq.~\eqref{eq:ME} preserves permutation symmetry, basis states $\ket{n_1,\ldots,n_d}$ are fully described by occupation numbers, where $n_\mu$ is the population of $\ket{g_\mu}$. These are typically entangled, as they consist of a symmetric superposition of excitations over $N$ particles. The population of the excited state is $n_e = N - \sum_{\mu} n_\mu$, with $n_e = 0$ in the final, steady state. Employing the ansatz $\rho = \sum_{\{n_\mu\}} P_{n_1,\ldots,n_d} \ket{n_1,\ldots,n_d}\bra{n_1,\ldots,n_d}$, Eq.~\eqref{eq:ME} reduces to a rate equation [see Supplementary Material (SM)~\cite{supp}]
\begin{equation}
\begin{split}
    \dot{P}_{n_1,\ldots,n_d} &= -\left(N - \sum_{\nu=1}^{d} n_\nu \right) \sum_{\mu=1}^{d} \Gamma_{\mu} (n_\mu + 1) P_{n_1, \ldots, n_\mu, \ldots, n_d} \\&+ \left(N - \sum_{\nu=1}^{d} n_\nu + 1 \right) \sum_{\mu=1}^{d} \Gamma_{\mu} n_\mu P_{n_1, \ldots, n_\mu-1,\ldots,n_d}
\end{split}
\label{eq:rate_eq}
\end{equation}
for the probabilities $P_{n_1,\ldots,n_d}$ of occupying the state $\ket{n_1,\ldots,n_d}$. This ansatz is justified even if there are coherences in the initial state (i.e., off-diagonal terms in the density matrix), since they are decoupled and hence do not affect population dynamics. While the rate equation holds for any permutationally symmetric initial state, below we choose the fully inverted state (i.e., $\ket{e}^{\otimes N}$). 

Although Eq.~\eqref{eq:rate_eq} can be efficiently simulated, it is non-trivial to obtain the steady state analytically. Moreover, the steady state is highly non-unique, since any combination of emitters in any ground state is a possible steady state. The trivial case of just one collective decay channel $(d=1)$ reduces to the problem of Dicke superradiance~\cite{dicke1954coherence,Gross1982superradiance}. We instead focus on the problem of multiple competing transitions.

We quantify \change{ground-state selection} by the dominant transition ratio (DTR), defined as the mean population density of the dominant ground state in the steady state, i.e.,
\begin{equation}
\text{DTR}=\frac{\overline{n_1}}{N},
\end{equation}
where $\overline{n_1}$ is the mean number of emitters in $\ket{g_1}$ as $t\to\infty$. For independent emitters, the marginal probability distribution $P(n_1) = \sum_{n_2,\ldots,n_d} P_{n_1,\ldots,n_d}$ for the dominant ground state is a binomial distribution whose average (normalized by $N$) is the DTR $= (1+\sum_{\mu>1} r_\mu)^{-1}$, determined solely by the decay ratios and independent of $N$ [see Fig.~\ref{fig:scheme}(c)]. Below, we prove that the DTR always converges to 1 as $N \to \infty$ for any number of collective decay channels, assuming $\Gamma_1 > \Gamma_{\mu>1}$. This effect can be attributed to the superradiant enhancement, which amplifies the dominant transition relative to all subdominant transitions. If $\Gamma_1 = \Gamma_2 = \ldots = \Gamma_d$, one obtains a uniform distribution for $P_{n_1,\ldots,n_d}$~\cite{orioli2022emergent}, as shown in Fig.~\ref{fig:scheme}(c).

To gain analytical insights from the rate equation, we transform it into a continuity equation of a fluid that flows in Euclidean space. In the limit of large $N$, we make the continuum approximation~\cite{degiorgio1971approximate} by setting $n_\mu \to x_\mu$, $\vec{x} = (x_1 \ldots x_d)^T \in \mathbb{R}_+^{d}$, and $f(n_1,\ldots,n_\mu,\ldots,n_d) - f(n_1,\ldots,n_\mu-1,\ldots,n_d) \to \frac{\partial}{\partial x_{\mu}} f(\vec{x})$
for an arbitrary differentiable function $f$. The rate equation is then approximated by the continuity equation
\begin{equation}
    \frac{\partial}{\partial t} P(\vec{x},t) = -\mathbf{\nabla} \cdot (\vec{v}(\vec{x}) P(\vec{x},t)),
\label{eq:fluid_PDE}
\end{equation}
which describes a fluid flow in $\mathbb{R}_+^d$, governed by the position-dependent velocity field $\vec{v}(\vec{x})$ with $\mu$-th component
\begin{equation}
    v_\mu(\vec{x}) = \Gamma_\mu \left(N - \sum_{\nu=1}^{d} x_\nu\right)(x_\mu + 1).
\label{eq:velocity_component}
\end{equation}
The flow comes to a stop as $\sum_\nu x_\nu \to N$, which physically corresponds to the system approaching its steady state, where no excited state population remains. The velocity field is illustrated for $d = 2$ in Fig.~\ref{fig:power_law}(a). The fully excited initial state corresponds to the origin of $\mathbb{R}_+^d$. Due to the non-uniqueness of the steady state, it is not sufficient to simply solve for $\partial_t P = 0$. 

Solving Eq.~\eqref{eq:fluid_PDE} analytically for arbitrary times is a formidable task. Instead, we employ a single-particle approximation where the fluid is idealized as a point particle initialized at the origin, with velocity dynamics $\text{d}\vec{x}/dt = \vec{v}(\vec{x})$. This reduces the partial differential equation in Eq.~\eqref{eq:fluid_PDE} to a coupled system of $d$ ordinary differential equations. Dividing the components of $\vec{v}(\vec{x})$, we readily find
\begin{equation}
    \frac{\text{d}x_\mu}{\text{d}x_\nu} = \frac{\Gamma_\mu}{\Gamma_\nu} \frac{x_\mu+1}{x_\nu + 1}
\end{equation}
for any pair of $\mu,\nu \in \{1,\ldots,d\}$. Integrating the above expression yields
\begin{equation}
    (x_\mu + 1)^{\Gamma_\nu} = (x_\nu+1)^{\Gamma_\mu}.
\label{eq:solution}
\end{equation}
The $d-1$ independent equations of the form~\eqref{eq:solution} define the particle trajectory in $\mathbb{R}_+^d$ starting from the origin. Generalizing Eq.~\eqref{eq:solution} to other permutationally symmetric initial states is straightforward (see SM~\cite{supp}). 

The steady state solution has a geometrical interpretation as the intersection between the particle trajectory and the hyperplane $\sum_{\mu=1}^{d} x_\mu = N$. To find an asymptotic solution, we assume that $x_\mu \gg 1$. The trajectory is then described by a simpler set of equations $x_\mu^{\Gamma_\nu} = x_\nu^{\Gamma_\mu}$. The constraint $\sum_{\mu=1}^{d} x_\mu = N$ can then be rewritten as $\sum_{\mu=1}^d x_d^{\Gamma_\mu/\Gamma_d} = N$. For simplicity, we consider the non-degenerate case where all $\Gamma_\mu$ are distinct (see SM~\cite{supp} for the degenerate case). Using the method of dominant balance~\cite{bender1999advanced} we obtain $x_d \approx N^{r_d}$, valid in the regime $1 \ll N^{r_d} \ll N$. In the regime $1 \ll N^{r_\mu} \ll N\; \forall\;\mu > 1$, an iterative method yields the self-consistent solution
\begin{equation}
    x_1 \approx N - \sum_{\mu=2}^d {N^{r_\mu}}, \quad x_\mu = N^{r_\mu}, \quad \mu > 1.
\label{eq:steadystate_approx}
\end{equation}
The dominant transition ratio thus reads
\begin{equation}
    \text{DTR} \approx \frac{x_1}{N} \approx 1 - \sum_{\mu=2}^{d} \frac{1}{N^{1-r_\mu}},
\label{eq:DTR_approx}
\end{equation}
which converges to unity as $N \to \infty$, with the slowest convergence characterized by the power law $\sim N^{r_2-1}$. Equation~\eqref{eq:DTR_approx} provides our first theoretical prediction for \change{ground-state selection}. Comparing with numerical simulations of the rate equation~\eqref{eq:rate_eq} for $d = 2$ in Fig.~\ref{fig:power_law}(b), the approximate formula~\eqref{eq:DTR_approx} agrees qualitatively, with higher accuracy attained for smaller $r_2$. 

Remarkably, we can solve for the DTR exactly in the asymptotic $N\rightarrow\infty$ limit. In the SM~\cite{supp}, we prove that the dynamics governed by the rate equation~\eqref{eq:rate_eq} is integrable, and derive the complete set of $N+1$ independent conserved quantities. We thus overcome the problem of non-unique steady states. However, because of the complexity of the exact solution, we only use it to compute the DTR, which reads
\begin{equation}
    \text{DTR} = \frac{\overline{n_1}}{N} = 1 - \sum_{\mu = 2}^{d} \frac{\tilde{\Gamma}(1-r_\mu)}{N^{1-r_\mu}},
\label{eq:DTR_conservation}
\end{equation}
where $\tilde{\Gamma}(\cdot)$ is the gamma function. The exact solution yields the same scaling as the approximate formula~\eqref{eq:DTR_approx}, and is in excellent agreement with numerical simulations, as shown in Fig.~\ref{fig:power_law}(b). This agreement explains why the approximation of Eq.~\eqref{eq:DTR_approx} becomes more accurate for smaller $r_\mu$, since $\tilde{\Gamma}(1-r_\mu) = 1 + O(r_\mu)$. The error in the prefactor of Eq.~\eqref{eq:DTR_approx} likely arises from the single-particle approximation.

Our formalism can also be easily extended to include non-collective decay channels, modeled as a leakage at a rate $\Gamma_{\text{leak}}$~\cite{supp}. Going back to the fluid model, this adds an extra component $v_{0} = \Gamma_{\text{leak}}\left(N - \sum_{\mu=1}^d x_\mu - x_{0} \right)$ to the velocity field and modifies $\sum_{\nu=1}^d x_\nu \to \sum_{\nu=0}^d x_\nu$ in Eq.~\eqref{eq:velocity_component}. A similar analysis for large $N$ yields $x_0 \approx (\Gamma_{\text{leak}}/\Gamma_1) \ln N$, in the same regime as the validity of Eq.~\eqref{eq:DTR_approx}. This justifies the omission of non-collective decay in our model, since $x_0 \ll x_\mu \sim N^{r_\mu}$, and the approximate DTR reads
\begin{equation}
    \text{DTR} \approx 1 - \sum_{\mu=2}^{d} \frac{1}{N^{1-r_\mu}} -\frac{\Gamma_{\text{leak}}}{\Gamma_1} \frac{\ln N}{N}.
\label{eq:DTR_continuum_leakage}
\end{equation}
In the non-competing scenario with only one collective decay channel $(d = 1)$, the DTR always converges to unity as $\sim \ln N / N$ (even if $\ket{e} \to \ket{g_1}$ is not the most dominant transition). This recovers the well-established result of Dicke superradiance in the presence of local decay~\cite{pupillo,malz2022large}. \change{Moreover, this implies that ground-state selection due to collective decay dominates when $\Gamma_{\text{leak}}/\Gamma_1 \ll N^{r_2}/\ln N$, which is realistic in experimental setups.}

The avalanche-like behavior of the dominant transition not only impacts the population of the dominant ground state, but also lowers the fluctuations of the probability distribution dramatically. For $d=2$, we numerically observe (for small $r_2$) that the relative fluctuation $\delta n_1/\overline{n_1}$ in the dominant ground state population vanishes as $N \to \infty$ faster than $\sim N^{-1/2}$ (expected from independent decay). This causes the steady state distribution to become sharply peaked at density $n_1/N = 1$, as indicated by the cumulative probability distribution in Fig.~\ref{fig:power_law}(c) and in the SM~\cite{supp}. By the Bhatia-Davis inequality~\cite{bhatia2000better}, the relative fluctuation can be bounded as $\delta n_1/\overline{n_1} \leq \sqrt{(1-\text{DTR})/\text{DTR}}$ which vanishes like $\sim N^{(r_2 - 1)/2}$. While this bound is not tight, it justifies the single-particle approximation made in our theoretical analysis.

Finally, the convergence time $T$ to the steady state can be estimated within the fluid model under the single-particle approximation. By noting that $\text{d}t = \text{d}x_1/v_1$, we find (see SM~\cite{supp})
\begin{equation}
   T = \frac{1}{\Gamma_1}\int_\mathcal{C} \frac{\text{d}x_1}{(N- x_1 - \sum_{\mu>1} x_\mu)(x_1+1)} \approx \frac{2-r_2}{\Gamma_1} \frac{\ln N}{N},
\label{eq:time_continuum_leakage}
\end{equation}
where $\mathcal{C}$ is the trajectory defined by Eq.~\eqref{eq:solution}. The presence of competing collective decay channels affects the well-known superradiant timescale of $\Gamma_1 T \sim \ln N / N$~\cite{Gross1982superradiance} only by a constant factor.

\begin{figure}
\centering
\subfloat{%
\includegraphics[width=0.95\linewidth]{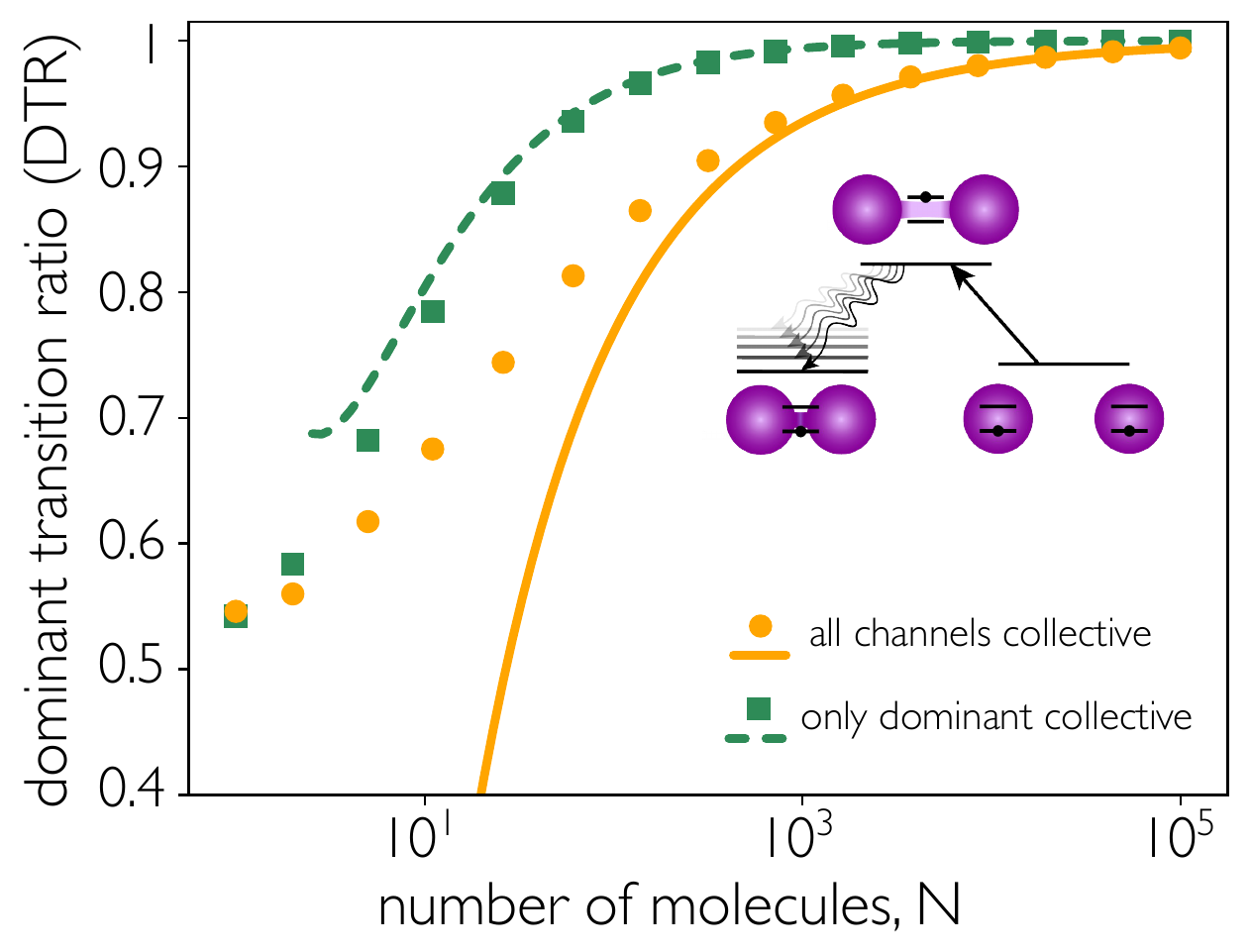}%
  \label{}%
}
\caption{Targeted photoassociation of a molecular dimer. Weakly bound molecules are created by optical excitation and spontaneously decay to various rovibrational ground-electronic states. The dominant transition ratio (DTR) is plotted against the initial number of weakly bound $\text{Sr}_2$ molecules in the $(1)1_u(\nu'=-1,J'=1)$ state. Decay is considered to four states, $(\nu=-1,J=0,2)$ and $(\nu=-2,J=0,2)$ with branching ratios 0.54, 0.27, 0.13, and 0.06, respectively. Points show the numerical solution from the Monte-Carlo simulation of the rate equation~\eqref{eq:rate_eq} for the non-competing scenario where only the dominant transition is collectively enhanced ($\square$) and the competing scenario with all four collective channels ($\circ$). The dashed line denotes the analytical prediction~\eqref{eq:DTR_continuum_leakage} for the non-competing scenario with $\Gamma_{\text{leak}}/\Gamma_1 = 0.852$. The solid line denotes the prediction~\eqref{eq:DTR_conservation} for the competing scenario. Predictions are valid for $N\gg1$.}
 \label{fig:molecule}
\end{figure} 

As a possible application of \change{ground-state selection}, we analyze the creation of ultracold diatomic molecules via photoassociation. In one-color photoassociation, laser-cooled atoms form weakly bound molecules that spontaneously decay into various more tightly-bound states~\cite{JonesRMP06}. The vibrational and rotational branching ratios are dictated by Franck-Condon factors and angular-momentum H\"{o}nl-London factors, respectively. Specifically, we examine ultracold strontium dimers that lack Feshbach resonances and must be produced optically~\cite{stellmer2012creation,reinaudi2012optical}. Strontium dimers have narrow optical transitions and a structureless ground state, making them well-suited for metrology~\cite{zelevinsky2008precision,leung2023terahertz,kondov2019molecular,ZelevinskyTiberi_arXiv_Sr2IsotopeShifts}. We consider 
photoassociation via the state $(1)1_u(\nu'=-1,J'=1)$~\cite{leung2023terahertz}, where $\nu'$ and $J'$ are the excited state vibrational and rotational quantum numbers, respectively, and negative vibrational numbers count down from the dissociation threshold. This state has a dominant decay path to the $\nu=-1, J=0$ ground state, with an overall branching ratio of $\sim54\%$ (molecular parameters are detailed in the SM~\cite{supp}). Our protocol uses a short, strong pulse to create $N$ weakly bound molecules from a dense sample of ultracold atoms. Alternatively, effective decay directly from unbound atoms into ground-state molecules can be engineered via a continuous off-resonant drive~\cite{inouye}. 

The avalanche-like behavior dramatically enhances the fraction of molecules in the dominant ground state, as shown in Fig.~\ref{fig:molecule}. If only one transition is collective, the dominant transition ratio rapidly approaches unity, as shown in Ref.~\cite{pupillo}. This can be realized by engineering the dielectric environment to be frequency-selective (either via a cavity~\cite{Kampschulte_2018} or a photonic crystal~\cite{perezrios}). We note that \change{ground-state selection} occurs in the weak coupling limit of cavity QED, distinguishing it from the so-called ``polaritonic chemistry'' in strong coupling~\cite{feist,hertzog}. Figure~\ref{fig:molecule} demonstrates that a nearly-pure sample of molecules is created even under maximal competition, where all decay channels are collective. Collective enhancement of all transitions can occur in broadband cavities (or for molecules with hyperfine structure, which yields smaller frequency differences between levels) or in single-mode fibers~\cite{Marquez-Mijares2023} within the ``mirror configuration''~\cite{polzik}. This effect can also occur in free space with dense, subwavelength molecular clouds. However, in this scenario, collisions can lead to significant losses. For polar molecules, collisions can be prevented via electric field~\cite{Valtolina20Dipolar} or microwave shielding~\cite{Anderegg21Observation,schindewolf,Lin23Microwave,Bigagli2024}. Alternatively, they can be suppressed by trapping molecules in optical tweezers to form ordered arrays~\cite{anderegg,cheuk}. As these systems are extended, their dynamics are not constrained to the permutationally symmetric subspace. Nevertheless, we expect \change{ground-state selection} to occur for molecules placed close to waveguides (at arbitrary distances) or arranged in ordered two- and three-dimensional arrays of subwavelength lattice constant~\cite{sierra, mok2024universal}, albeit with reduced scaling.

In summary, collective decay holds promise for a wide range of quantum systems, from closing open transitions in atoms~\cite{sutherland2017superradiance,masson2024dicke} and preventing parasitic decay in solid-state emitters~\cite{sipahigil2016integrated,scheibner2007superradiance,lange2024superradiant} to directing emission into dielectric nanostructures instead of unwanted modes~\cite{liedl2024observation,cardenas-lopez2023many}. While we have primarily focused on an initially inverted ensemble, our treatment captures \change{ground-state selection} from a large range of multiply excited states. Future research directions include studying the potential for many-body-enhanced metrology~\cite{ding2022enhanced}, due to the sensitivity of the ground state populations to the decay ratios. Other interesting avenues include the observation of symmetry breaking, which can occur if multiple dominant transitions have identical decay rates, akin to mirror symmetry breaking predicted in waveguide QED~\cite{cardenas-lopez2023many}.

\textit{Acknowledgements.---} We thank Kon Leung,  Luis A. Orozco, and Maximilian Schemmer for helpful discussions. We acknowledge support by the National Science Foundation through the CAREER Award (No. 2047380) and the QLCI program (grant No. OMA-2016245), the Air Force Office of Scientific Research through their Young Investigator Prize (grant No. 21RT0751) and through grant No. FA9550-1910328, ARO through the MURI program (Grant No. W911NF-20-1-0136), DARPA (Grant No. W911NF2010090), as well as by the David and Lucile Packard Foundation and the Brown Science Foundation. The Institute for Quantum Information and Matter is an NSF Physics Frontiers Center.

\bibliography{bib}

\end{document}


\title{Supplementary Material: Ground-state selection via many-body superradiant decay}

\author{Wai-Keong Mok}
\affiliation{Institute for Quantum Information and Matter, California Institute of Technology, Pasadena, CA 91125, USA} 
\author{Stuart J. Masson}
\affiliation{Department of Physics, Columbia University, New York, New York 10027, USA}
\author{Dan M. Stamper-Kurn}
\affiliation{Department of Physics, University of California, Berkeley, California 94720}
\affiliation{Challenge Institute for Quantum Computation, University of California, Berkeley, California 94720}
\affiliation{Materials Sciences Division, Lawrence Berkeley National Laboratory, Berkeley, California 94720}
\author{Tanya Zelevinsky}
\affiliation{Department of Physics, Columbia University, New York, New York 10027, USA}
\author{Ana Asenjo-Garcia}
\affiliation{Department of Physics, Columbia University, New York, New York 10027, USA}

\maketitle

\tableofcontents

\section{Collective jump operators}

Here, we provide a simple derivation for the collective jump operators $\hat{A}_\mu$ corresponding to the transition $\ket{e}\to\ket{g_\mu}$. In the permutation-symmetric subspace, we label the basis states as $\ket{n_1,n_2,\ldots,n_d}$, where $n_\mu$ denotes the number of emitters in the ground state $\ket{g_\mu}$. In terms of the product basis kets,
\begin{equation}
    \ket{n_1,n_2,\ldots,n_d} = \begin{pmatrix}
        N \\ n_1\, n_2 \ldots n_d \, n_e
    \end{pmatrix}^{-1/2} \left( \ket{\underbrace{g_1 \ldots g_1}_{n_1} \ldots \underbrace{g_d \ldots g_d}_{n_d} \underbrace{e \ldots e}_{n_e}} + \text{permutations} \right),
\label{eq:permbasis}
\end{equation}
where $n_e = N - \sum_{\mu=1}^{d} n_\mu$ and $N\choose{n_1 \ldots n_e}$ is the multinomial coefficient. The action of the collective jump operators on $\ket{n_1,n_2,\ldots,n_d}$ reads,
\begin{equation}
\begin{split}
    \hat{A}_\mu \ket{n_1,n_2,\ldots,n_\mu,\ldots n_d} &= \begin{pmatrix}
        N \\ n_1\, n_2 \ldots n_d\, n_e
    \end{pmatrix}^{-1/2} \hat{A}_\mu \left( \ket{\underbrace{g_1 \ldots g_1}_{n_1} \ldots \underbrace{g_d \ldots g_d}_{n_d} \underbrace{e \ldots e}_{n_e}} + \text{permutations} \right)  \\&\propto \ket{n_1,n_2,\ldots,n_\mu+1,\ldots n_d}.
\end{split}
\end{equation}
The proportionality constant can be worked out via a simple counting argument. Firstly, $\hat{A}_\mu$ acting on a product basis ket in Eq.~\eqref{eq:permbasis} produces a superposition of $n_e$ product basis kets. We can thus write
\begin{equation}
\begin{split}
    \hat{A}_\mu \ket{n_1,n_2,\ldots,n_\mu,\ldots n_d} &= \begin{pmatrix}
        N \\ n_1\, n_2 \ldots n_d\, n_e
    \end{pmatrix}^{-1/2} \left[\text{sum of} \quad\begin{pmatrix}
        N \\ n_1\, n_2 \ldots n_d\, n_e
    \end{pmatrix} n_e \quad \text{product basis kets}  \right].
\end{split}
\end{equation}
However, this is overcounting since there are only $N! / (n_1! \ldots (n_\mu + 1)! \ldots n_d! (n_e-1)!)$ unique product basis kets. Since the state is permutationally-symmetric, each unique product basis ket is repeated the same number of times in the sum. Hence,
\begin{equation}
\begin{split}
    \hat{A}_\mu \ket{n_1,n_2,\ldots,n_\mu,\ldots n_d} &= \begin{pmatrix}
        N \\ n_1\, n_2 \ldots n_d\, n_e
    \end{pmatrix}^{-1/2} \times \frac{\begin{pmatrix}
        N \\ n_1\, n_2 \ldots n_d\,  n_e
    \end{pmatrix} n_e}{\begin{pmatrix}
        N \\ n_1\, n_2 \ldots (n_\mu+1) \ldots n_d (n_e-1)
    \end{pmatrix}} \\&\times  \left( \ket{\underbrace{g_1 \ldots g_1}_{n_1} \ldots \underbrace{g_\mu \ldots g_\mu}_{n_\mu+1}\underbrace{g_d \ldots g_d}_{n_d} \underbrace{e \ldots e}_{n_e-1}} + \text{permutations} \right).
\end{split}
\end{equation}
Since
\begin{equation}
\ket{n_1,n_2,\ldots,n_\mu+1,\ldots n_d} = \begin{pmatrix}
        N \\ n_1\, n_2 \ldots (n_\mu+1) \ldots n_d (n_e-1)
    \end{pmatrix}^{-1/2} \left( \ket{\underbrace{g_1 \ldots g_1}_{n_1} \ldots \underbrace{g_\mu \ldots g_\mu}_{n_\mu+1}\underbrace{g_d \ldots g_d}_{n_d} \underbrace{e \ldots e}_{n_e-1}} + \text{permutations} \right),
\end{equation}
we have
\begin{equation}
\begin{split}
    \hat{A}_\mu \ket{n_1,n_2,\ldots,n_\mu,\ldots n_d} &= \begin{pmatrix}
        N \\ n_1\, n_2 \ldots n_d\,n_e
    \end{pmatrix}^{-1/2} \times \frac{\begin{pmatrix}
        N \\ n_1\, n_2 \ldots n_d\, n_e
    \end{pmatrix} n_e}{\begin{pmatrix}
        N \\ n_1\, n_2 \ldots (n_\mu+1) \ldots n_d \, (n_e-1)
    \end{pmatrix}} \\&\times \begin{pmatrix}
        N \\ n_1\, n_2 \ldots (n_\mu+1) \ldots n_d\, (n_e-1)
    \end{pmatrix}^{1/2} \ket{n_1,n_2,\ldots,n_\mu+1,\ldots n_d},
\end{split}
\end{equation}
which can be simplfied as
\begin{equation}
    \hat{A}_\mu \ket{n_1,n_2,\ldots,n_\mu,\ldots n_d} = \sqrt{\left(N - \sum_{\nu=1}^{d} n_\nu\right)(n_\mu + 1)} \ket{n_1,n_2,\ldots,n_\mu+1,\ldots n_d}.
\label{eq:alpha_create}
\end{equation}
Correspondingly,
\begin{equation}
    \hat{A}_\mu^\dag \ket{n_1,n_2,\ldots,n_\mu,\ldots n_d} = \sqrt{\left(N - \sum_{\nu=1}^{d} n_\nu + 1\right)n_\mu} \ket{n_1,n_2,\ldots,n_\mu-1,\ldots n_d}.
\label{eq:alpha_destroy}
\end{equation}

\section{Derivation of the rate equation}

From the master equation
\begin{equation}
    \dot{\rho} = -i\left[\sum_{\mu=1}^{d} \chi_\mu \hat{A}_\mu^\dag \hat{A}_\mu,\rho \right] + \sum_{\mu=1}^{d} \Gamma_\mu \mathcal{D}[\hat{A}_\mu]\rho,
\label{eq:collectiveME}
\end{equation}
we derive the corresponding rate equation for the populations $P_{n_1,\ldots,n_d}$. Let us expand the density matrix in the permutation symmetric subspace as
\begin{equation}
    \rho = \sum_{\mu=1}^{d} \sum_{n_\mu, n_\mu^\prime=0}^{N} \rho_{n_1,\ldots,n_d}^{n_1^\prime,\ldots,n_d^\prime} \ket{n_1,\ldots,n_d}\bra{n_1^\prime,\ldots,n_d^\prime},
\end{equation}
with the basis states defined in Eq.~\eqref{eq:permbasis}. The diagonal terms correspond to the populations
\begin{equation}
    P_{n_1,\ldots,n_d} \equiv \rho_{n_1,\ldots,n_d}^{n_1,\ldots,n_d}
\end{equation}
of occupying the state $\ket{n_1,\ldots,n_d}$, while the off-diagonal terms are the coherences. Using the collective operators in Eqs.~\eqref{eq:alpha_create} and~\eqref{eq:alpha_destroy}, the master equation can be written as
\begin{equation}
\begin{split}
\partial_t \rho_{n_1,\ldots,n_d}^{n_1^\prime,\ldots,n_d^\prime} &= -i\sum_{\mu=1}^d \chi_\mu \rho_{n_1,\ldots,n_d}^{n_1^\prime,\ldots,n_d^\prime} \left[ \braket{n_1,\ldots,n_d|\hat{A}_\mu^\dag \hat{A}_\mu|n_1,\ldots,n_d} - \braket{n_1^\prime,\ldots,n_d^\prime|\hat{A}_\mu^\dag \hat{A}_\mu|n_1^\prime,\ldots,n_d^\prime} \right] \\&+ \sum_{\mu=1}^d \Gamma_\mu \Bigg\{ \rho_{n_1,\ldots,n_\mu-1,\ldots,n_d}^{n_1^\prime,\ldots,n_\mu^\prime-1,\ldots,n_d^\prime} \braket{n_1,\ldots,n_d|\hat{A}_\mu|n_1,\ldots,n_\mu-1,\ldots,n_d}\braket{n_1^\prime,\ldots,n_\mu^\prime-1,\ldots,n_d^\prime|\hat{A}_\mu^\dag|n_1^\prime,\ldots,n_d^\prime} \\&-\frac{1}{2} \rho_{n_1,\ldots,n_d}^{n_1^\prime,\ldots,n_d^\prime} \left[ \braket{n_1,\ldots,n_d|\hat{A}_\mu^\dag \hat{A}_\mu|n_1,\ldots,n_d} + \braket{n_1^\prime,\ldots,n_d^\prime|\hat{A}_\mu^\dag \hat{A}_\mu|n_1^\prime,\ldots,n_d^\prime} \right] \Bigg\}.  
\end{split}   
\end{equation}
From the definitions~\eqref{eq:alpha_create} and~\eqref{eq:alpha_destroy} of the collective jump operators, $\braket{n_1,\ldots,n_d|\hat{A}_\mu^\dag \hat{A}_\mu|n_1,\ldots,n_d} = (N - \sum_\nu n_\nu) (n_\mu + 1)$ and $\braket{n_1,\ldots,n_d|\hat{A}_\mu|n_1,\ldots,n_\mu-1,\ldots,n_d} = \sqrt{(N-\sum_\nu n_\nu +1)\,n_\mu}$. The master equation is thus

\begin{equation}
\begin{split}
\partial_t \rho_{n_1,\ldots,n_d}^{n_1^\prime,\ldots,n_d^\prime} &= -i\sum_{\mu=1}^{d} \chi_\mu \left[ \left(N-\sum_\nu n_\nu\right)(n_\mu+1) - \left(N-\sum_\nu n_\nu^\prime\right)(n_\mu^\prime+1) \right]\rho_{n_1,\ldots,n_d}^{n_1^\prime,\ldots,n_d^\prime} \\&+ \sum_{\mu=1}^{d} \Gamma_\mu \Bigg\{ \sqrt{\left(N-\sum_\nu n_\nu+1\right)n_\mu\left(N-\sum_\nu n_\nu^\prime+1\right)n_\mu^\prime}\, \,\rho_{n_1,\ldots,n_\mu-1,\ldots,n_d}^{n_1^\prime,\ldots,n_\mu^\prime-1,\ldots,n_d^\prime} \\&- \frac{1}{2} \left[\left(N-\sum_\nu n_\nu\right)(n_\mu+1)+\left(N-\sum_\nu n_\nu^\prime\right)(n_\mu^\prime+1)\right] \rho_{n_1,\ldots,n_d}^{n_1^\prime,\ldots,n_d^\prime} \Bigg\}.    
\end{split}
\end{equation}
Note that the populations are decoupled from the coherences, which decay to zero at long times. Thus, for the purposes of obtaining the steady state populations, we can ignore the dynamics of the coherences. The diagonal terms from setting $n_\mu^\prime = n_\mu$ read
\begin{equation}
    \dot{P}_{n_1,\ldots,n_d} = \sum_{\mu=1}^d \Gamma_\mu \left[\left(N - \sum_\nu n_\nu + 1\right) n_\mu P_{n_1,\ldots,n_\mu-1,n_d} - \left(N - \sum_\nu n_\nu\right)(n_\mu+1) P_{n_1,\ldots,n_d}  \right],
\end{equation}
which leads to the rate equation for the populations.

\section{Effect of non-collective decay channels}

Non-collective decay channels can be modelled as a leakage out of the subspace spanned by $\ket{e}$ and the $d$ ground states $\ket{g_1},\ldots,\ket{g_d}$. We model non-collective decay channels by introducing an additional ground state $\ket{g_0}$ for each emitter. The transition $\ket{e} \to \ket{g_0}$ occurs at a rate of $\Gamma_{\text{leak}}$ which is not collectively enhanced. Defining the lowering operator $\hat{\sigma}_i^- = (\ket{g_0}\bra{e})_i$, the non-collective decay contributes a dissipative term $\Gamma_{\text{leak}} \sum_i \mathcal{D}[\hat{\sigma}_i^-]\rho$ to the master equation. We label states by $\ket{n_0;n_1,\ldots,n_d}$, where $n_\nu$ is the number of emitters in the ground state $\ket{g_\nu}$, $\nu = 0,1,\ldots,d$. The number of emitters in the excited state is thus $n_e = N - \sum_{\nu=0}^d n_\nu$. However, due to the presence of independent decay, the states $\ket{n_0;n_1,\ldots,n_d}$ cannot be interpreted as a symmetrized state, since the local jump operators $\hat{\sigma}_i$ do not preserve permutation symmetry, although the density matrix remains symmetric if $\Gamma_{\text{leak}}$ is the same for all $i\in\{1,\ldots,N\}$~\cite{shammah2018open}. Instead, we modify the notation to use $\ket{n_0;n_1,\ldots,n_d}$ as a shorthand to denote any state containing $n_\mu$ emitters in the ground state $\ket{g_\mu}$, with $\mu = \{0,\ldots,d\}$.

Consider the state $\ket{n_0;n_1,\ldots,n_d}$. The probability of a collective transition with the operator $\hat{A}_\mu$ is
\begin{equation}
    p_\mu \propto \Gamma_\mu \braket{n_0;n_1,\ldots,n_d|\hat{A}_\mu^\dag \hat{A}_\mu|n_0;n_1,\ldots,n_d} = \Gamma_\mu \left(N - \sum_{\nu=0}^d n_\nu \right) (n_\mu + 1),
\end{equation}
and the total leakage probability is
\begin{equation}
    p_\text{leak} \propto \Gamma_{\text{leak}}\sum_{i=1}^N \braket{n_0;n_1,\ldots,n_d| \hat{\sigma}_i^+\hat{\sigma}_i^-|n_0;n_1,\ldots,n_d} = \Gamma_{\text{leak}} \left(N - \sum_{\nu=0}^d n_\nu \right).
\end{equation}
The point to note here is that the collective transition probabilities and the total leakage probability depend only on the occupation numbers $\{n_\mu\}$ and hold even if $\ket{n_0;n_1,\ldots,n_d}$ is not a permutationally symmetric state. Thus, for our purposes of calculating the DTR, the exact form of $\ket{n_0;n_1,\ldots,n_d}$ is not relevant.

In the continuum approximation, this is modelled as a fluid flow in $\mathbb{R}_+^{d+1}$, with the velocity vector field $\vec{v}(\vec{x}) = (v_0,v_1,\ldots,v_d)^T$ having components
\begin{equation}
    v_0(\vec{x}) = \Gamma_{\text{leak}}\left(N - \sum_{\nu=0}^d x_\nu\right),
\end{equation}
and
\begin{equation}
    v_\mu(\vec{x}) = \Gamma_\mu \left(N - \sum_{\nu=0}^d x_\nu\right) (x_\mu + 1), \quad \mu = 1,\ldots,d.
\end{equation}
Dividing the components of $\vec{v}(\vec{x})$, we have (for $\mu = 1,\ldots,d$)
\begin{equation}
    \frac{\text{d}x_0}{\text{d}x_\mu} = \frac{\Gamma_{\text{leak}}}{\Gamma_\mu} \frac{1}{x_\mu + 1},
\end{equation}
which can be integrated to obtain
\begin{equation}
    x_0 = \frac{\Gamma_{\text{leak}}}{\Gamma_\mu} \ln(x_\mu + 1).
\end{equation}
In the steady state, $x_0 + \sum_{\mu=1}^d x_\mu = N$, so
\begin{equation}
    x_0 + \sum_{\mu=1}^d \left( e^{\Gamma_{\mu} x_0 /\Gamma_{\text{leak}}} - 1 \right) = N.
\label{eq:continuum_leak}
\end{equation}
Solving this for $x_0$ exactly is difficult. Instead, for large $N$, we use dominant balance to get
\begin{equation}
    e^{\Gamma_1 x_0 /\Gamma_{\text{leak}}} \approx N \implies x_0 \approx \frac{\Gamma_{\text{leak}}}{\Gamma_1} \ln N.
\end{equation}
It remains to check that the omitted terms in Eq.~\eqref{eq:continuum_leak} are subdominant to $N$. The largest omitted term is
\begin{equation}
    e^{\Gamma_2 x_0 / \Gamma_{\text{leak}}} \approx e^{r_2 \ln N} = N^{r_2} \ll N
\end{equation}
for a sufficiently large $N$. This is the same regime of validity as for our approximate DTR formula. Thus, for large $N$, the various $x_\mu \sim N^{r_\mu}$ scale as a power law with $N$, while $x_0 \sim \ln N$ only scales logarithmically with $N$. We thus conclude that non-collective decay does not affect the asymptotic behavior of our approximate DTR.

In the special case of $d = 1$, the system reduces to the well-studied problem of Dicke superradiance in two-level systems, with additional leakage that is not collectively enhanced. There is no competition between the various collective decay channels. As a result, the mean population density of the ground state $\ket{g_1}$ scales as
\begin{equation}
    \frac{\overline{n_1}}{N} \approx \frac{x_1}{N} \sim 1 - \frac{\Gamma_{\text{leak}}}{\Gamma_1} \frac{\ln N}{N}
\end{equation}
which always converges to unity, even if $\Gamma_1 < \Gamma_{\text{leak}}$. 

\subsection{Independent decay channels in the non-interacting limit}

When all the decay channels are non-collective, the master equation describing the independent decay of the emitters can be derived by taking the non-interacting limit of the collective master equation~\eqref{eq:collectiveME}. To derive this, notice that the collective master equation can be rewritten as
\begin{equation}
    \dot{\rho} = \sum_{\mu = 1}^{d} \sum_{i,j=1}^{N} \Gamma_\mu^{ij} \left(\hat{\sigma}_{\mu,i} \rho \hat{\sigma}_{\mu,j}^\dagger - \frac{1}{2} \hat{\sigma}_{\mu,j}^\dagger \hat{\sigma}_{\mu,i} \rho - \frac{1}{2} \rho \hat{\sigma}_{\mu,j}^\dagger \hat{\sigma}_{\mu,i}\right), 
\end{equation}
with
\begin{equation}
    \Gamma_{\mu}^{ij} = \Gamma_\mu \quad \forall i,j = 1,\ldots,N
\label{eq:Gammaij}
\end{equation}
and
\begin{equation}
    \hat{\sigma}_{\mu,i} = (\ket{g_\mu}\bra{e})_i
\end{equation}
is the lowering operator for the $i$-th emitter corresponding to the transition type $\mu$. This describes the microscopic dissipative interactions with interaction strengths $\Gamma_{\mu}^{ij}$ between emitters $i$ and $j$. The non-interacting limit is defined by setting the interaction strength $\Gamma_\mu^{ij} = 0$ for all $i \neq j$, such that $\Gamma_{\mu}^{ij} = \Gamma_\mu \delta^{ij}$, which yields the master equation
\begin{equation}
    \dot{\rho} = \sum_{i=1}^{N} \sum_{\mu = 1}^{d} \Gamma_\mu  \mathcal{D}[\hat{\sigma}_{\mu,i}]\rho.
\label{eq:independent}
\end{equation}
In the steady state when all the emitters have decayed, the probability distribution for the ground state is simply given by the multinomial distribution
\begin{equation}
    P_{n_1,\ldots,n_d} = {N \choose n_1 n_2 \ldots n_d} \prod_{\mu=1}^d \left(\frac{\Gamma_\mu}{\sum_{\nu=1}^d \Gamma_\nu}\right)^{n_\mu}.
\end{equation}
The dominant transition ratio can be easily evaluated as
\begin{equation}
    \text{DTR} = \frac{\Gamma_1}{\Gamma_1 + \sum_{\mu>1}\Gamma_\mu} = \frac{1}{1 + \sum_{\mu>1} r_\mu}
\end{equation}
independent of $N$.
\section{Degenerate case of identical decay rates}
In the main text, we analyze the ground-state selection in the non-degenerate case, i.e, $\Gamma_1 > \Gamma_2 > \ldots > \Gamma_d$. Here, we analyze the degenerate case where some of the $\Gamma_\mu$ are identical. If all $\Gamma_\mu$ are the same, the steady state probability distribution $P_{n_1,\ldots,n_d}$ is the uniform distribution supported on points $(n_1,\ldots,n_d)$ satisfying the constraint $n_1 + \ldots + n_d = N$ ~\cite{orioli2022emergent}. More generally, we consider $d$ ground states split into $L$ degenerate blocks, with degeneracies $\{m_1,m_2,\ldots,m_L\}$ and $\sum_{\mu=1}^{L} m_\mu = d$. The decay rates are ordered as $\Gamma_1 > \Gamma_2 > \ldots > \Gamma_L$ with the corresponding ratios $1 > r_2 > \ldots > r_L$. Note that the degeneracy here refers to identical decay rates (e.g., $\Gamma_1 = \Gamma_2 = \ldots = \Gamma_{m_1}$) and not identical ground state energies.

Following the continuum and single-particle approximation explained in the main text, we have
\begin{equation}
    x_\mu^{\Gamma_\nu} \approx x_\nu^{\Gamma_\mu},
\label{eq:gs_reln}
\end{equation}
where $x_\mu$ is interpreted as the population of any of $m_\mu$ ground states in the $\mu$-th degenerate block. Then, $x_\mu \approx x_L^{r_\mu/r_L}$. The steady state condition is written as
\begin{equation}
    \sum_{\mu=1}^L m_\mu x_\mu = N \implies \sum_{\mu=1}^{L} m_\mu x_L^{r_\mu/r_L} \approx N.
\end{equation}
Using dominant balance, we get
\begin{equation}
    x_L \approx \left(\frac{N}{m_1}\right)^{r_L}.
\end{equation}
Equation~\eqref{eq:gs_reln} then gives the other ground state populations as
\begin{equation}
    x_\mu \approx \left(\frac{N}{m_1}\right)^{r_\mu}, \quad \mu = 2,\ldots,L.
\end{equation}
Plugging this into the steady state condition,
\begin{equation}
    m_1 x_1 + \sum_{\mu=2}^{L} m_\mu \left(\frac{N}{m_1}\right)^{r_\mu} \approx N.
\end{equation}
Defining the dominant transition ratio (DTR) as $x_1/N$, we obtain
\begin{equation}
    \text{DTR} \approx \frac{1}{m_1} - \sum_{\mu=2}^{L} \frac{m_\mu}{m_1^{1+r_\mu}N^{1-r_\mu}},
\end{equation}
which recovers the result in the main text when $m_1 = m_2 = \ldots = m_L = 1$. The factor of $1/m_1$ in the first term reflects the fact that as $N \to \infty$, the population density is shared equally between the $m_1$ dominant ground states with the largest decay rate $\Gamma_1$.

\section{Fluctuations of the ground state population density}

We numerically simulate the rate equation for $d =2$ ground states. From the steady state probability distribution of the dominant ground state $P(n_1)$, we find the mean and variance of the distribution, to compute the relative fluctuation $\delta n_1 / \overline{n_1}$, where $\delta n_1$ is the standard deviation for the population of $n_1$. We then fit the relative fluctuation to the phenomenological power law $\sim N^{-\beta}$ for $N$ up to $10^4$, to extract the fluctuation power-law index $\beta$. The fit is done for small values of $r_2 = \Gamma_2/\Gamma_1$, since larger values of $r_2$ require a larger $N$ to accurately capture the asymptotic behavior. From our fit, we estimate $\beta = 0.930 - 0.842 r_2$. In the main text, we show that the variance $\text{Var}(n_1/N) \leq (1-\text{DTR})\text{DTR}$. Thus,
\begin{equation}
    \frac{\delta n_1}{\overline{n_1}} = \frac{\sqrt{\text{Var}(n_1/N)}}{\text{DTR}} \leq \sqrt{\frac{1-\text{DTR}}{\text{DTR}}} \sim N^{(-1+r_2)/2}
\end{equation}
for $d=2$. Comparing this with the asymptotic behavior $\delta n_1/\overline{n_1} \sim N^{-\beta}$, this gives the lower bound
\begin{equation}
    \beta \geq \frac{1}{2}(1-r_2).
\end{equation}
Our estimated $\beta$ from the fitting satisfies this bound for all $r_2 \in (0,1)$.

\begin{figure}
\centering
\subfloat{%
\includegraphics[width=0.65\linewidth]{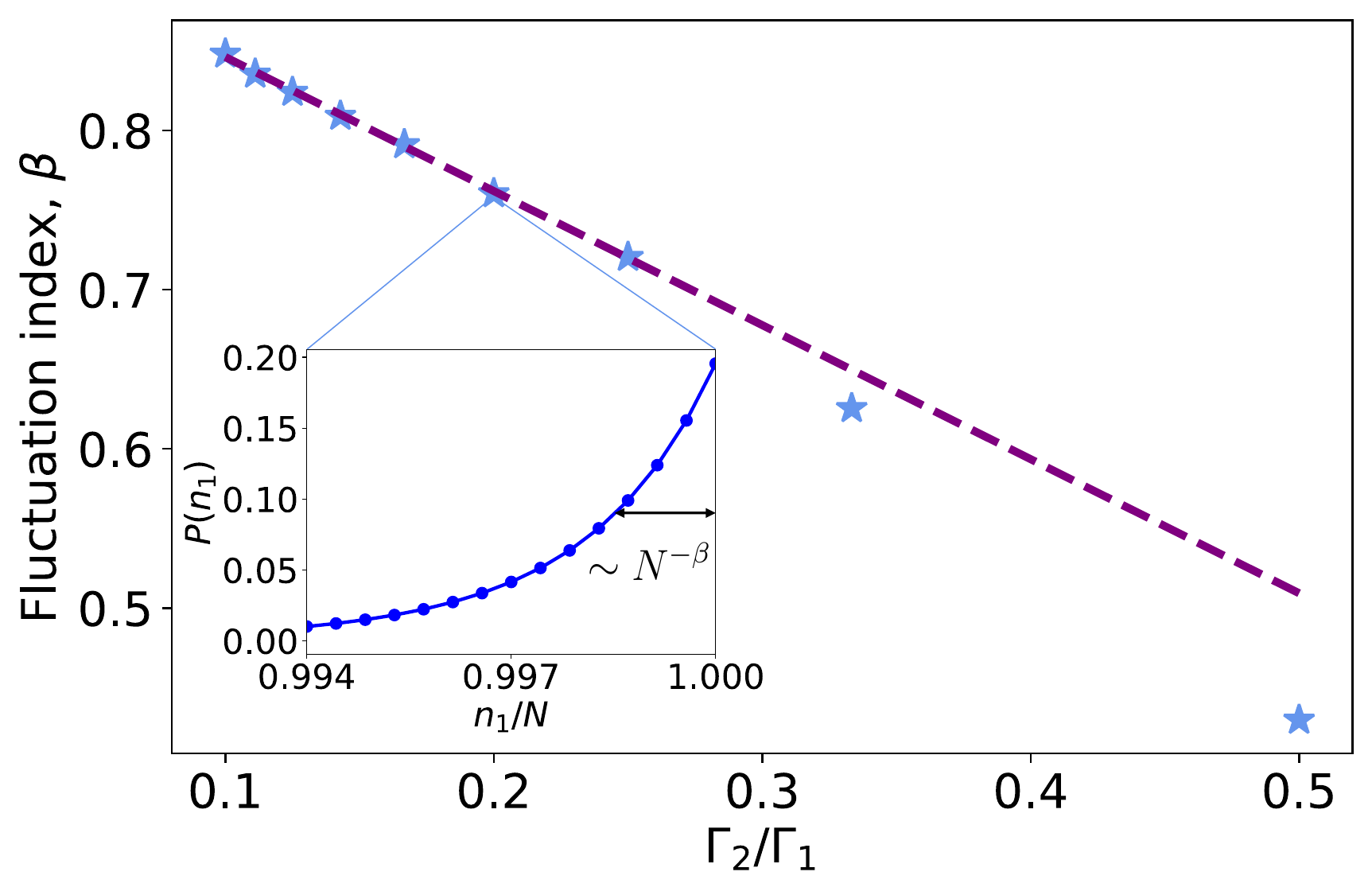}%
  \label{}%
}
\caption{Power law index $\beta$ from fitting the tail of the relative fluctuation $\delta n_1 / \overline{n_1}$ of the dominant ground state population to $\sim N^{-\beta}$. The dashed line indicates the phenomenological fit $\beta = 0.930-0.842 r_2$. The inset shows the population distribution of the dominant ground state for $r_2 = 0.2$, plotted between $0.994 \leq n_1/N \leq 1$, with a width of order $N^{-\beta}$.}
 \label{fig:fluct}
\end{figure}

\section{Estimate of the convergence time to steady state}
From the continuum approximation, we have the velocity field $\vec{v}(\vec{x})$ with components
\begin{equation}
    v_\mu \equiv \frac{\text{d}x_\mu}{\text{d}t} = \left(N - \sum_{\mu=1}^d x_\mu \right)(x_\mu + 1).
\end{equation}
In the single-particle approximation, the trajectory $\mathcal{C}$ in $\mathbb{R}_+^d$ is defined by the set of equations
\begin{equation}
    (x_\mu + 1)^{\Gamma_\nu} = (x_\nu + 1)^{\Gamma_\mu}
\end{equation}
for all pairs $\mu,\nu \in \{1,\ldots,d\}$, in the region $\sum_{\mu=1}^d x_\mu \leq N$. The convergence time can be written as a line integral
\begin{equation}
    T = \int_\mathcal{C} \frac{\text{d} x_1}{v_1} = \frac{1}{\Gamma_1}\int_\mathcal{C} \frac{\text{d}x_1}{(N-x_1 - \sum_{\mu>1}x_\mu)(x_1+1)}
\end{equation}
from $x_1 = 0$ to $x_1 = N - \sum_{\mu>1} N^{r_\mu}$. As in the main text, we consider the non-degenerate case of $r_\mu$ being all different for simplicity. Performing the integral directly is not trivial. Instead, we can obtain an approximation by splitting the trajectory $\mathcal{C}$ into two parts~\cite{malz2022large}: $\mathcal{C}_1$, where the particle travels along $\mathcal{C}$ from $x_1 = 0$ to $x_1 = N^a$ for some $0<a<1$, and $\mathcal{C}_2$, where the particle travels along $\mathcal{C}$ from $x_1 = N^a$ to $x_1 = N - \sum_{\mu>1} N^{r_\mu}$. There is no loss of generality in setting the intermediate point to be at $x_1 = N^a$, since we can always choose a point on $\mathcal{C}$ with $1 \ll x_1 \ll N$ and define $a$ appropriately. 

For the trajectory $\mathcal{C}_1$, $N \gg \{x_1, x_{\mu>1}\}$, so
\begin{equation}
    T_1 \equiv \int_{\mathcal{C}_1} \frac{\text{d}x_1}{v_1} \approx \frac{1}{N \Gamma_1} \int_0^{N^a} \frac{dx_1}{x_1+1} \approx \frac{a}{\Gamma_1} \frac{\ln N}{N}.
\end{equation}

For the trajectory $\mathcal{C}_2$, $x_1 \gg x_{\mu>1}$, so
\begin{equation}
\begin{split}
    T_2 \equiv \int_{\mathcal{C}_2} \frac{\text{d}x_1}{v_1} &\approx \frac{1}{\Gamma_1} \int_{N^a}^{N-\sum_{\mu>1}N^{r_\mu}} \frac{\text{d}x_1}{(N-x_1)(x_1+1)} \approx \frac{1}{N\Gamma_1} \ln \left(\frac{x_1+1}{N-x_1}\right) \bigg|_{N^a}^{N-\sum_{\mu>1}N^{r_\mu}} \\&\approx \frac{1}{N\Gamma_1} \ln \left(\frac{N^2}{N^{r_2+a}}\right) = \frac{2-r_2-a}{\Gamma_1} \frac{\ln N}{N}.
\end{split}
\end{equation}
This gives the convergence time $T$ as
\begin{equation}
    T = T_1 + T_2 \approx \frac{2-r_2}{\Gamma_1} \frac{\ln N}{N}.
\end{equation}
The final result for $T$ does not depend on $a$, which is consistent since the intermediate point is chosen arbitrarily. Thus, adding competing collective decay channels does not affect the convergence timescale of $\sim \ln N / N$, but only changes the prefactor.

\section{Integrability and conservation laws of the rate equation}

Since the steady state is non-unique and depends on the initial conditions, the rate equation encodes conservation laws that we now derive. Our goal is to find a function $h(n_1,\ldots,n_d)$ that satisfies the conservation law $\partial_t \overline{h(n_1,\ldots,n_d)} \equiv \sum_{n_1,\ldots,n_d} h(n_1,\ldots,n_d) \dot{P}_{n_1,\ldots,n_d} = 0$, where $\dot{P}_{n_1,\ldots,n_2}$ is defined by the rate equation. Explicitly,
\begin{equation}
\begin{split}
    0 &= \sum_{n_1,\ldots,n_d} \sum_{\mu=1}^d \Gamma_\mu \left[-\left(N - \sum_\nu n_\nu\right) \left(n_\mu + 1\right)P_{n_1,\ldots,n_\mu,\ldots,n_d} + \left(N-\sum_\nu n_\nu + 1\right) n_\mu P_{n_1,\ldots,n_\mu-1,\ldots,n_d} \right] h(n_1,\ldots,n_d) \\ &= \sum_{n_1,\ldots,n_d} \sum_{\mu=1}^{d} \Gamma_\mu \left(N - \sum_\nu n_\nu \right) (n_\mu + 1) \left[ h(n_1,\ldots,n_\mu+1,\ldots,n_d) - h(n_1,\ldots,n_\mu,\ldots,n_d) \right] P_{n_1,\ldots,n_d}.
\end{split}
\label{eq:difference_eq}
\end{equation}
Subsituting the separable ansatz
\begin{equation}
    h(n_1,\ldots,n_d) = \prod_{\mu = 1}^{d} h_\mu(n_\mu)
\end{equation}
into the conservation law, and dividing through by $h(n_1,\ldots,n_d)$, we have
\begin{equation}
    \sum_\mu \Gamma_\mu (n_\mu+1) \left( \frac{h_\mu(n_\mu + 1)}{h_\mu(n_\mu)} - 1 \right) = 0,
\end{equation}
using the fact that the probabilities $P_{n_1,\ldots,n_d}$ are arbitrary. Since each term in the sum above only depends on $n_\mu$ that are arbitrary, we must have  
\begin{equation}
    \Gamma_\mu (n_\mu+1) \left( \frac{h_\mu(n_\mu + 1)}{h_\mu(n_\mu)} - 1 \right) = c_\mu
\label{eq:recurrence}
\end{equation}
where $c_\mu$ are constants satisfying the constraint
\begin{equation}
    \sum_{\mu=1}^d c_\mu = 0.
\end{equation}
Equation~\eqref{eq:recurrence} is a recurrence relation for $h_\mu(n_\mu)$, which has the general solution
\begin{equation}
    h_\mu(n_\mu) \propto \frac{\tilde{\Gamma}(1+n_\mu+c_\mu/\Gamma_\mu)}{n_\mu!},
\end{equation}
where $\tilde{\Gamma}(z)$ is the gamma function. Multiplying the $h_\mu$'s together and choosing the normalization factor such that $\overline{h} = 1$ for the fully excited state (described by the probability distribution $P_{n_1,\ldots,n_d} = \prod_\mu \delta_{n_\mu,0}$), we obtain the generating function
\begin{equation}
    h(n_1,\ldots,n_d) = \prod_{\mu=1}^d \frac{\tilde{\Gamma}(1+n_\mu+c_\mu/\Gamma_\mu)}{n_\mu! \tilde{\Gamma}(1+c_\mu/\Gamma_\mu)}.
\label{eq:genfunction}
\end{equation}
By construction, the steady state distribution satisfies $\overline{h} = 1$ with $\sum_\mu n_\mu = N$. The generating function $h(n_1,\ldots,n_d)$ is a polynomial of degree $n_\mu$ in the variable $c_\mu$, which can be seen by expanding out Eq.~\eqref{eq:genfunction} using the definition $\tilde{\Gamma}(1+z) = z\tilde{\Gamma}(z)$ of the gamma function iteratively,
\begin{equation}
    h(n_1,\ldots,n_d) = \prod_{\mu=1}^d \frac{1}{n_\mu !} \prod_{j=1}^{n_\mu} \left(j + \frac{c_\mu}{\Gamma_\mu}\right).
\end{equation}
This expression holds for $n_\mu > 0$. If $n_\mu = 0$, the product $\prod_j (j + c_\mu/\Gamma_\mu)$ is replaced by $1$. Thus, $h(n_1,\ldots,n_d)$ encodes $N+1$ independent conserved quantities, which are the coefficients of the polynomial. A trivial conserved quantity can be obtained by setting all $c_\mu = 0$,
\begin{equation}
    1 = \overline{h} = \sum_{n_1,\ldots,n_d} P_{n_1,\ldots,n_d},
\end{equation}
which is simply the conservation of probability. By choosing suitable values of $c_\mu$, we seek non-trivial conserved quantities.

\subsection{Dominant transition ratio}
Let us choose $c_\mu = \Gamma_\mu$ for some particular $\mu > 1$, and all other $c_{\nu > 1} = 0$. From the constraint $\sum_{k=1}^{d} c_k = 0$, this fixes $c_1 = - \Gamma_\mu$. Substituting this into the conservation law $\overline{h} = 1$, we obtain
\begin{equation}
    \bigg\langle \frac{\tilde{\Gamma}(1+n_1 -r_\mu)\tilde{\Gamma}(2+n_\mu)}{n_1! n_\mu!}\bigg\rangle = \tilde{\Gamma}(1- r_\mu).
\end{equation}
The average on the LHS is taken with respect to the steady state probability distribution. We denote $r_\mu = \Gamma_\mu/\Gamma_1 < 1$. Assuming that the main contribution from $n_\mu$ comes from $1 \ll n_\mu \ll N$, we use Stirling's approximation to get
\begin{equation}
    \overline{n_1^{-r_\mu} n_\mu} \approx \tilde{\Gamma}(1- r_\mu).
\end{equation}
Since $r_\mu < 1$, the gamma function is always finite and positive. Substituting the steady state constraint $n_1 = N - \sum_{\nu > 1} n_\nu$, and Taylor expanding in powers of $1/N$, we have
\begin{equation}
    \overline{n_\mu} + \frac{r_\mu}{N} \sum_{\nu>1} \overline{n_\mu n_\nu} \approx \tilde{\Gamma}(1- r_\mu) N^{r_\mu}.
\label{eq:taylorexpanded}
\end{equation}
To proceed further, we assume $\sum_{\nu > 1} \overline{n_\mu n_\nu} \ll N \overline{n_\mu}$ to finally arrive at
\begin{equation}
    \overline{n_\mu} \approx \tilde{\Gamma}(1- r_\mu) N^{r_\mu}
\label{eq:first_moment}
\end{equation}
which gives the DTR as
\begin{equation}
    \text{DTR} = \frac{\overline{n_1}}{N} = 1 - \frac{1}{N}\sum_{\mu > 1} \overline{n_\mu} = 1 - \sum_{\mu>1} \frac{\tilde{\Gamma}(1-r_\mu)}{N^{1-r_\mu}}.
\end{equation}
The assumption $\sum_{\nu > 1} \overline{n_\mu n_\nu} \ll N \overline{n_\mu}$ can be thought of as a mean-field approximation $\overline{n_\mu n_\nu} \approx \overline{n_\mu}\,\,\overline{n_\nu}$. Then, using Eq.~\eqref{eq:first_moment}, we have $\sum_{\nu > 1} \overline{n_\mu n_\nu} \approx \overline{n_\mu} \sum_{\nu>1} \overline{n_\nu} \approx \overline{n_\mu} \sum_{\nu>1} N^{r_\nu} \ll N \overline{n_\mu} $ which is self-consistent. Since the population density $n_1/N \in [0,1]$ is bounded, we can use the Bhatia-Davis inequality~\cite{bhatia2000better} to bound the variance by
\begin{equation}
    \text{Var}(n_1/N) \leq (1-\text{DTR})\text{DTR} \sim \sum_{\mu>1} \frac{\tilde{\Gamma}(1-r_\mu)}{N^{1-r_\mu}}.
\end{equation}

Although Eq.~\eqref{eq:first_moment} is finite for any $r_\mu < 1$, one has to be careful about the divergence of the gamma function as $r_\mu \to 1^-$. Since $\tilde{\Gamma}(z) \sim 1/z$ near $z = 0$, we get
\begin{equation}
    \overline{n_\mu} \approx \frac{N^{r_\mu}}{1-r_\mu}, \quad r_\mu \to 1^-.
\end{equation}
To observe ground-state selection, we demand $\overline{n_\mu} \ll N$ which gives the condition
\begin{equation}
    N \gg y^y, \quad y = \frac{1}{1-r_\mu}.
\end{equation}
Physically, this means that for $r_\mu$ close to $1$, we require a larger system size to suppress the sub-dominant collective transitions. As an estimate, if $r_\mu = 0.9$, then our theoretical predictions hold for $N \gg 10^{10}$.

\subsection{Continuum approximation}
We can also derive the conservation laws in the continuum approximation. Applying the continuum approximation to the difference equation~\eqref{eq:difference_eq}, we have
\begin{equation}
    \sum_{\mu = 1}^{d} \Gamma_{\mu} (x_\mu + 1) \frac{\partial}{\partial x_\mu} h(\vec{x}) = 0.
\end{equation}
Using the separable ansatz
\begin{equation}
    h(\vec{x}) = \prod_{\mu=1}^{d} h_\mu (x_\mu),
\end{equation}
we obtain
\begin{equation}
    \sum_{\mu=1}^{d} \Gamma_\mu (x_\mu+1) \frac{h_\mu^\prime(x_\mu)}{h_\mu(x_\mu)} = 0.
\end{equation}
Each term in the sum depends on an independent coordinate $x_\mu$, so the terms in the sum must all be equal to some constant $c_\mu$
\begin{equation}
    \Gamma_\mu (x_\mu+1) \frac{h_\mu^\prime(x_\mu)}{h_\mu(x_\mu)} = c_\mu
\end{equation}
with the constraint $\sum_{\mu=1}^{d} c_\mu = 0$. Solving these equations, we get the generating function
\begin{equation}
    h(\vec{x}) = \prod_{\mu=1}^{d} (x_\mu+1)^{c_\mu/\Gamma_\mu}.
\label{eq:continuum_genfun}
\end{equation}
The fully excited initial state is represented by the Dirac delta distribution $P(\vec{x}) = \delta^d(\vec{x})$, which satisfies $\overline{h(\vec{x})} = 1$. Equation~\eqref{eq:continuum_genfun} is the continuous analog of Eq.~\eqref{eq:genfunction}. Similar to the discrete case, we can choose $c_\mu = \Gamma_\mu$ for some $\mu > 1$ and all other $c_{\nu > 1} = 0$, which constraints $c_1 = -\Gamma_\mu$. Assuming that the main contribution comes from $1 \ll x_\mu \ll N$ and using $x_1 = N - \sum_{\mu>1} x_\mu$, we have
\begin{equation}
    \overline{x_\mu} \approx N^{r_\mu},
\end{equation}
which gives the approximate DTR as
\begin{equation}
    \text{DTR} \approx 1 - \sum_{\mu>1} \frac{1}{N^{1-r_\mu}}.
\end{equation}
This agrees with the approximate formula derived in the main text using the single-particle approximation.

\newpage
\section{Molecular parameters for photoassociation}

\begin{table*}[h!]
 \begin{center}
 \begin{tabular}{| c | c | c | c |}
 \hline
 Vibrational quantum number $\nu$ & Rotational quantum number $J$ & Branching factor & Binding energy, MHz$/h$\\\hline
 -1 & 0 & 0.54 & 137\\\hline
 -1 & 2 & 0.27 & 67 \\\hline
 -2 & 0 & 0.13 & 1400 \\\hline
 -2 & 2 & 0.06 & 1227\\\hline
 \end{tabular}
 \end{center}
 \caption{Predominant decay paths from the excited $(1)1_u(\nu'=-1,J'=1)$ state of a strontium dimer~\cite{BorkowskiPrivate}. There are four states with significant branching ratios, those with rotational quantum number $J=\{0,2\}$ and vibrational quantum number $\nu=\{-1,-2\}$ (where negative vibrational numbers count down from the dissociation threshold) \cite{MajewskaThesis}. All other states have branching ratios of $<0.01$ and are ignored in the calculations. The minimum separation between these states is for the two $\nu=-1$ states, where the separation between the $J=0$ and $J=2$ states is $~70$ MHz~\cite{LeungThesis}, significantly larger than the linewidth of the excited state which is $\sim15$ kHz~\cite{ZelevinskyPRL06}.\label{table:molecularnumbers}}
 \end{table*}

\bibliography{suppbib}